\documentclass[aps,pre,preprint]{revtex4-1}

\usepackage[utf8]{inputenc}
\usepackage{amsmath,amssymb,latexsym,mathtools,natbib,graphicx}

\mathtoolsset{showonlyrefs=true}

\begin{document}


\def\[{\begin{equation}}
\def\]{\end{equation}}

\def\rr{\ensuremath\rho}
\def\Ell{\ensuremath{\mathcal L}}
\def\reals{\ensuremath{\mathbb R}}
\def\bigo{\ensuremath{\mathcal O}}
\def\ints{\ensuremath{\mathbb Z}}
\def\del{\ensuremath{\vec\nabla}}
\def\rec{\ensuremath{\mathrm{rec}}}
\def\dd{\ensuremath d}
\def\stripe{\ensuremath{\mathrm{stripe}}}
\def\double{\ensuremath{\mathrm{double}}}
\def\forked{\ensuremath{\mathrm{forked}}}

\newcommand\norm[1]{\ensuremath{\left\|#1\right\|}}
\newcommand\abs[1]{\ensuremath{|#1|}}
\newcommand\Abs[1]{\ensuremath{\left|#1\right|}}
\newcommand\bound[1]{\ensuremath{\partial #1}}
\newcommand\unit[1]{\ensuremath{\hat{\vec{#1}}}}
\newcommand\pd[2]{\ensuremath{\frac{\partial#1}{\partial#2}}}

\renewcommand\vec[1]{\ensuremath{\boldsymbol{\mathbf{#1}}}}


\title{Energy Driven Pattern Formation in Planar Dipole--Dipole Systems in
the Presence of Weak Noise}
\author{Jaron Kent-Dobias}
\author{Andrew J.~Bernoff}
\affiliation{Harvey Mudd College}
\date{\today}

\begin{abstract}
	We study pattern formation in planar fluid systems driven by intermolecular
	cohesion (which manifests as a line tension) and dipole--dipole repulsion
	which are observed in physical systems including ferrofluids in Hele-Shaw
	cells and Langmuir layers.  When the dipolar repulsion is sufficiently
	strong, domains undergo forked branching reminiscent of viscous fingering.
	A known difficulty with these models is that the energy associated with
	dipole--dipole interactions is singular at small distances.  Following
	previous work, we demonstrate how to ameliorate this singularity and show
	that in the macroscopic limit only the scale of the microscopic
	details relative to the macroscopic extent of a system is relevant and develop an expression for the system
	energy that depends only on a generalized line tension, $\Lambda$, that in
	turn depends logarithmically on that scale.  We conduct numerical studies
	that use energy minimization to find equilibrium states.  Following the
	subcritical bifurcations from the circle, we find a few highly symmetric
	stable shapes, but nothing that resembles the observed diversity of
	experimental and dynamically simulated domains.  The application of a weak
	random background to the energy landscape stabilizes a sm\"org\r asbord of
	domain morphologies recovering the diversity observed experimentally.  With
	this technique, we generate a large sample of qualitatively realistic shapes
	and use them to create an empirical model for extracting $\Lambda$ with high accuracy using
	only a shape's perimeter and morphology .   
\end{abstract}

\maketitle

\section{Introduction}

\begin{figure}
	\centering
	\includegraphics{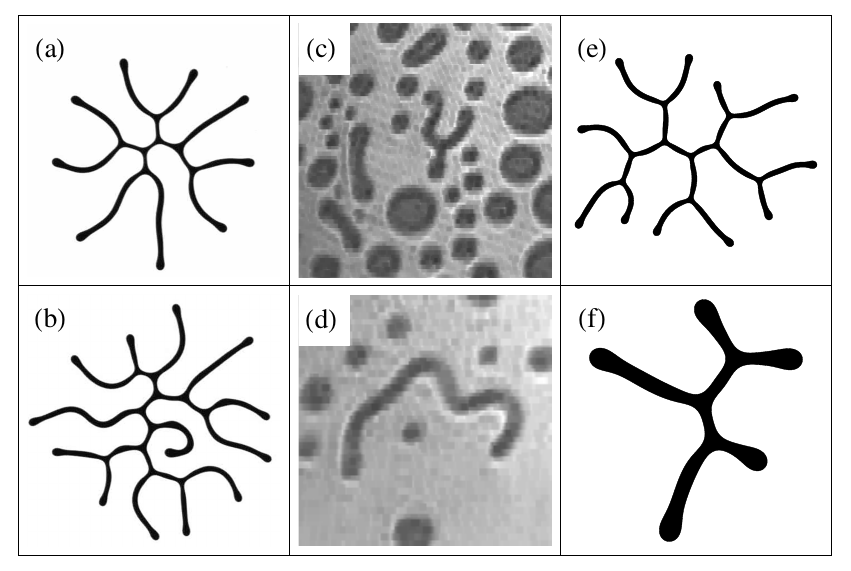}
	\caption{Examples of two-dimensional dipole-mediated systems in experiments.
		(a,~b) Ferrofluid enclosed in a Hele-Shaw cell.  Images provided by
		D.~P.~Jackson \cite{GJ94,DEGJL93,LGJ92}. (c,~d) 8--CB Langmuir films, or
		monolayers of polymer molecules, condensed into their fluid phase.  Images
		provided by E.~K.~Mann \cite{MP99}. (e, f) Results of our numeric
		simulations.}
	\label{figs:ferrofluid}
\end{figure}

A wide variety of two-dimensional systems driven by competition between
strong, short-ranged attractive and long-ranged, dipole-like repulsive forces
exhibit striking phenomenological similarities.  This interplay leads to the
formation of intricate and tree-like structures.  Substantial work has been
done in characterizing the physics, dynamics, and morphology of these systems
\cite{C80,CM80a,CM80b,CM80c,CZ83,MM88,VM90,M90,MK92,LGJ92,DEGJL93,GJ94,JGC94,KM94,KJM95,LG95,SA95,MK96,CD95,CD99,MP99,KF02,EDCFB98,O98,CD99,KF02,HHF04,JH07,J08}.  Langmuir monolayers
\cite{VM90,KM94,LG95,MK96,MP99,KF02} and ferrofluid confined to a Hele-Shaw cell
\cite{C80,CM80a,CM80b,CM80c,CZ83,VM90,LGJ92,DEGJL93,JGC94,GJ94,KM94,LG95,MK96,CD95,EDCFB98,CD99,KF02,JH07,J08} are of particular interest, and shape
formation and stability in these systems have been studied extensively in
experiment (see examples in Fig.~\ref{figs:ferrofluid}(a--d)).

The inherent complexity of dipole-mediated systems has also inspired numerical
simulations using dynamic evolution of some particular system's equations of
motion \cite{ABMMWZ07,DEGJL93,HHF04,JGC94,LG95,T08,CD99,CD95,CZ83,EDCFB98}.  For example, Cebers et
al.\ \cite{CD99,CM80a} and Jackson et
al.\ \cite{LGJ92,DEGJL93} considered Hele-Shaw systems, providing analytic and asymptotic
expressions for the energy and stability of a variety of domain structures, including
 circles and rectangles.  McConnell et al.\ built an effective
theoretical formalism for describing the energy of Langmuir domains, and
determined analytically the stability of circular and stripe domains to
harmonic perturbations \cite{VM90,M90,MK92,KJM95}.  McConnell et al.\ were
able to show that the detailed physical parameters that describe Langmuir
systems can be reduced to a single parameter \cite{KM93}.  Though this
reduction greatly simplifies the state space of these systems, it does not
appear to have been used by other researchers after McConnell.  The energy
formalism that we will introduce here, though different from the one used by
McConnell et al., proceeds along nearly identical lines to reduce the
parameter space to one dimension.  We use this energy formalism numerically to
realize static equilibrium states via energy extremization. These methods can
resolve subcritical branches for harmonic bifurcations associated with the
hysteresis discovered by Cebers et al.\ \cite{CZ83} and later studied by Jackson and his
collaborators \cite{JH07,J08}.  In the
absence of noise, we discovered stable domains are characterized by a few,
highly symmetric morphologies.  We argue that the rich qualitative structure
seen in dynamic studies and experiment is due to the presence of random
imperfections modelled as variations in the energy landscape, and show that a
simple empirical rule exists for accurately determining the state parameter of
dipole-mediated systems.

\section{Analysis}

Consider a compact region $\Omega\subset\reals^2$ which describes the spatial
extent of a dipole-mediated domain.  The energy of such a domain is given by
\cite{M90,K14}
\[
	E=\alpha A+\lambda\ell
	+\frac{\mu^2}2\iint\limits_\Omega\iint\limits_\Omega
	\frac{g(\norm{\vec r-\vec r'})}{\norm{\vec r-\vec r'}^3}\,\dd{A'}\,\dd A
	+\iint\limits_\Omega V(\vec r)\,\dd A.
	\label{energy.area}
\]
The first two terms are proportional to the area $A$ and the perimeter $\ell$
of $\Omega$ respectively, where $\lambda$ is the line tension.  Since we only consider
domains constrained to have constant area, the first term is irrelevant to
system behavior and will henceforth be neglected.  The third term is the
energy due to the dipole--dipole interaction, a continuum approximation of
mutually interacting dipole pairs.  The fourth term is the energy due to an
arbitrary static external potential.  The constant $\mu$ is an effective
dipole density, and the function $g(r)$ is the pair correlation function for
the system, which gives the probability distribution that a dipole is
displaced from another by $r$.  In disordered systems, like the ones we
consider, $g$ must be isotropic (and therefore radially symmetric), have
$g(0)=0$ (as particles cannot exist atop each other), and $g(r)$ must be well
approximated by $1$ if $r>\Delta$ for some interparticle length scale $\Delta$
\cite{H90}.  In order for \eqref{energy.area} to converge, $g(r)$ must vanish
at least as quickly as $r^2$ as $r$ tends to zero.  For different physical
systems $g(r)$ can take on a variety of forms, all of which are highly
dependent on the microscopic details of the particular system.  As we will see
presently, the particular form of $g(r)$ is unimportant to the behavior of the
domain when the microscopic parameter $\Delta$ is much smaller than the
characteristic length scale of that domain, e.g., $\ell$.

Using Green's theorem, we may convert \eqref{energy.area} to a line integral
over the domain's boundary $\bound\Omega$ in an analogous fashion to that done
by McConnell et al.\ \cite{MM88,MK92}.  In this case, we have for the energy
\[
	E=\lambda\ell-\frac{\mu^2}2\oint\limits_{\bound\Omega}\oint\limits_{\bound\Omega}
	\Phi(\norm{\vec r-\vec r'})(\unit n\cdot\unit n')\,\dd{s'}\,\dd s
	+\oint\limits_{\bound\Omega}\vec\Psi(\vec r)\cdot\unit n\,\dd s.
	\label{energy.line}
\]
Here, $\unit n$ is the unit normal to the parameterization $\vec s$, $\Phi(r)$
is such that $\del^2\Phi(r)=g(r)r^{-3}$, and $\vec\Psi(\vec r)$ is such that
$\del\cdot\vec\Psi=V$.  We find via direct integration of the Laplacian that
\begin{align}
	\Phi(r)
	&=\int_r^\infty \frac1{r'}\int_{r'}^{\infty}r''\left[\frac{g(r'')}{{r''}^3}\right]\,\dd{r''}\,\dd{r'}\\
	&=\frac{g(r)}r+\int_r^\infty \left[\frac{g'(r')}{r'} + \frac1{r'}\int_{r'}^{\infty}
	\frac{g'(r'')}{r''}\,\dd{r''}\right]\,\dd{r'},
%
	\label{phi}
\end{align}
where we have integrated by parts twice to reach the final expression.  We note that $\Phi(r)$ may have a logarithmic singularity at $r=0$ which is harmless as it is integrable. We would like to simplify this expression by considering the limit of small
$\Delta$.  Let the double integration in \eqref{energy.line} be represented by
\[
	I\equiv\frac12\oint\limits_{\bound\Omega}\oint\limits_{\bound\Omega}
	\Phi(\norm{\vec r-\vec r'})(\unit n\cdot\unit n')\,\dd{s'}\,\dd s.
\]
We may now explicitly parameterize the line integral by arc length, yielding
\[
	I=\frac12\int_0^\ell\int_0^\ell
	\Phi(\norm{\vec r(s)-\vec r(s')})[\unit n(s)\cdot\unit n(s')]\,\dd s'\,\dd s.
\]
Defining $\sigma\equiv s'-s$, we now reparameterize the integral to the form
\[
	I=\frac12\int_0^\ell\int_{-\frac\ell2}^{\frac\ell2}
	\Phi(\norm{\vec r(s)-\vec r(s+\sigma)})[\unit n(s)\cdot\unit n(s+\sigma)]
	\,\dd\sigma\,\dd s.
	\label{int.reparam}
\]
Consider some function $j(r,\Delta)$ with the following two properties:
\[\lim_{\Delta\to0}j(r,\Delta)=\frac1r
	\quad\text{and}\quad
	J(\Delta)\equiv\frac12\int_{-\frac\ell2}^{\frac\ell2}j(\abs\sigma,\Delta)
	\,\dd\sigma<\infty.
	\label{jint}
\]
Adding and subtracting the same quantity involving $j(r,\Delta)$ from
\eqref{int.reparam} yields
\[
	I=\frac12\int_0^\ell\int_{-\frac\ell2}^{\frac\ell2}
	\left\{\Phi(\norm{\vec r(s)-\vec r(s+\sigma)})[\unit n(s)\cdot\unit
	n(s+\sigma)]-j(\abs\sigma,\Delta)\right\}\,\dd\sigma\,\dd s
	+J(\Delta)\ell.
	\label{int.final}
\]
Now take the limit as $\Delta\to0$ in the integrand of \eqref{int.final}.  The
function $j(r,\Delta)$ behaves as described in \eqref{jint}.  Since, as
$\Delta\to0$, $g(r)$ tends to unity for all $r\in(0,\infty)$, it follows that
$g'(r)\simeq0$ in this range as well, and \eqref{phi} yields
\[
	\lim_{\Delta\to0}\Phi(r)=\frac1r.
\]
Carrying this limit through within the integral, we find
\[
	I\simeq\frac12\int_0^\ell\int_{-\frac\ell2}^{\frac\ell2}
	\left[\frac{\unit n(s)\cdot\unit n(s+\sigma)}{\norm{\vec r(s)
	-\vec r(s+\sigma)}}-\frac1{\abs\sigma}\right]\,\dd\sigma\,\dd s+J(\Delta)\ell.
	\label{i.approx}
\]
This integral, which without the addition of $j(r,\Delta)$ would be singular,
now converges.  This can be seen by examining the behavior of the integrand as
$\sigma\to0$, or
\[
	\frac{\unit n(s)\cdot\unit n(s+\sigma)}{\norm{\vec r(s)-\vec r(s+\sigma)}}
	-\frac1{\abs\sigma}
	=\frac{1+\bigo(\sigma^2)}{\abs\sigma+\bigo(\sigma^3)}-\frac1{\abs\sigma}
	=\bigo(\sigma^2).
\]
We have been able to completely remove the dependence on $g(r)$ from the
integration.  This may seem worrisome, since $g(r)$ implicitly contained
information about the microscopic parameters of the system, like the length
scale $\Delta$.  This parameter still enters the energy, but now through the
function $J(\Delta)$, which we have yet to choose.  If we pick
$j(r,\Delta)=[\Theta(r-\Delta/2)+\Theta(-r-\Delta/2)]/r$, where $\Theta$ is
the Heaviside function, it follows immediately from \eqref{jint} that
$J(\Delta)=\log\frac\ell\Delta$, and we have
\[
	I=\frac12\int_0^\ell\int_{-\frac\ell2}^{\frac\ell2}
	\left[\frac{\unit n(s)\cdot\unit n(s+\sigma)}{\norm{\vec r(s)
	-\vec r(s+\sigma)}}-\frac1{\abs\sigma}\right]+\ell\log\frac\ell\Delta.
\]
This choice of $j(r,\Delta)$ is motivated mostly by its simplicity.  Many
other options are available, though for consistency with the small-$\Delta$
approximation one usually must then expand $J(\Delta)$ about
$\frac\Delta\ell=0$ and use the highest order term.  In any such case, given
the asymptotic behavior of $j(r,\Delta)$ as defined above, the highest order
term will be proportional to $\log\frac\ell\Delta$, and the particular choice
of $j$ will only modify the proportionality constant.  The error due to taking
this macroscopic limit in \eqref{i.approx} goes as $\Delta^3$ \cite{K14}.

We are now able to write a more explicit form of the energy function
\eqref{energy.line},
\[
	E=\lambda\ell-
	\frac{\mu^2}2\oint\limits_{\bound\Omega}\int_{-\frac\ell2}^\frac\ell2
	\left[\frac{\unit n(s)\cdot\unit n(s+\sigma)}{\norm{\vec r(s)-\vec r(s+\sigma)}}-
	\frac1{\abs\sigma}\right]\,\dd\sigma\,\dd s-\mu^2\ell
	\log\frac\ell\Delta+\oint\limits_{\bound\Omega}\vec\Psi(\vec r)\cdot\unit
	n\,\dd s.
	\label{line.energy}
\]
To fully describe a system we are modelling, one must also enforce that the
area of the domain is constant, or
\begin{align}
	A=\iint\limits_\Omega\dd A
	=\frac12\oint\limits_{\bound\Omega}\unit z\cdot
	\bigg(\vec r\times\pd{\vec r}s\bigg)\;\dd s.
	\label{cont.area.cost.dim}
\end{align}
Note that, in the absence of the last term describing an auxiliary field,
\eqref{line.energy} is precisely what was found by McConnell et al.\
\cite{KM93}.  However, in that study, the expression is derived for a
particular $g(r)$, while we have now shown that \emph{any} typical $g(r)$ will
lead to a system described by the same energy.  It will be convenient to
non-dimensionalize this system for ease of analysis and numerics.  First,
define $R\equiv\sqrt{A/\pi}$, the characteristic radius of the domain.  Then
define
\begin{align}
	F\equiv\frac E{\mu^2R}, &&
	L\equiv\frac\ell R, &&
	\Lambda\equiv\frac{\lambda}{\mu^2}-\log\frac R\Delta, &&
	\vec\rr\equiv\frac{\vec r}R, &&
	\vec\Pi\equiv\frac{\vec\Psi}{\mu^2R}.
\end{align}
Upon substitution of these quantities into \eqref{line.energy} and
simplification, the nondimensional energy is
\[
	F
	=\Lambda L-
	\frac12\oint\limits_{\bound\Omega}\int_{-\frac L2}^\frac L2
	\left[\frac{\unit n(s)\cdot\unit n(s+\sigma)}{\norm{\vec\rr(s)-\vec\rr(s+\sigma)}}
	-\frac1{\abs\sigma}\right]\,\dd\sigma\,\dd s
	-L\log L+\oint\limits_{\bound\Omega}\vec\Pi(s)\cdot\unit n(s)\,\dd s.
	\label{energy.final}
\]
When nondimensionalized, the area constraint \eqref{cont.area.cost.dim} becomes
\begin{align}
	\pi=\frac12\oint\limits_{\bound\Omega}\unit z\cdot
	\bigg(\vec\rho\times\pd{\vec\rho}s\bigg)\;\dd s.
	\label{cont.area.cost}
\end{align}
These expressions only depend on a single parameter, $\Lambda$, and on the shape of
the domain $\Omega$.  The parameter $\Lambda$ can be interpreted physically as
as an effective line tension, normalized by the dipole density $\mu^2$ and
shifted by the log of the ratio of the microscopic and macroscopic length
scales of the system.  An interesting but perhaps unintuitive feature of this
is that the instabilities we observe occur when $\Lambda$ is negative, a
situation physically obtainable due to the shift.

In the limit of large $\Lambda$, the energy minimization problem is dominated
by perimeter minimization and circular domains are the stable minimizer in
this regime.  The value of $\Lambda$ at which circular domains become unstable
is of interest because it marks the transition from this simple regime to one
characterized by more interesting structure.  Setting $\vec\rr(s)=\unit x\cos
s+\unit y\sin s$, one can use \eqref{energy.final} to determine the energy of
a circular domain explicitly, yielding
\[
	F_\circ(\Lambda)=2\pi(\Lambda+2-\log8)
\]
matching the known result from McConnell et al.\ and equivalent to
similar calculations by Cebers and his collaborators \cite{MM88,CM80a,CM80b}.  Define
$\Lambda_n$ as the critical value of $\Lambda$ at which a circular domain
becomes unstable to $n$th order sinusoidal perturbations of the type
$\delta\vec\rr_n(\theta)=\epsilon\cos(n\theta)\vec\rr(\theta)$.  These
critical values of $\Lambda$ are given by $\Lambda_n=\log8-Z_n$, where the
first few $Z_n$ are tabulated in \cite{M90} and an explicit form is given in
\cite{GJ94}; details of our calculation can be found in \cite{K14}.  We use
these critical instabilities to verify the accuracy of our numeric
simulations.

Another important previous result is the calculation of the  energy of a
rectangular domain, which was computed previously by McConnell et al.\
\cite{MM88} and Langer et al.\ \cite{LGJ92}.  If $a$ is the aspect ratio of a
rectangular domain, then the $x$ and $y$ dimensions of that domain are
$d_x=\sqrt{a\pi}$ and $d_y=\sqrt{\frac\pi a}$, respectively.  In the limit of
large $a$, or high aspect ratio, we find
\[
	F_\rec
	=2\sqrt{\pi a}\left(\Lambda-\frac12\log\frac\pi a\right)+\bigo(a^{-1/2}).
\]
The value of the aspect ratio at which the above energy is minimized is
$a(\Lambda)=\pi e^{-2(\Lambda+1)}$.  This corresponds to a domain perimeter of
$L_\rec(\Lambda)=2\pi e^{-\Lambda-1}$, and a rectangle energy of
$F_\rec(\Lambda)=-2\pi e^{-\Lambda-1}$.  Here, we find that as $\Lambda$
decreases, the aspect ratio of rectangular domains grows exponentially, and as
a result so do their perimeters.  We also find that the minimum energy of a
rectangle increases exponentially with decreasing $\Lambda$.  Because the
energy of a rectangular domain decreases so quickly, it becomes lower than
that of a circular domain when $F_\rec(\Lambda)=F_\circ(\Lambda)$.  This
transition happens at $\Lambda\simeq-1.374$.  Comparing this with
$\Lambda_2=-1.254$, the point at which the circle first becomes unstable, we
see that the transition to lower rectangle energy may be related to the
transition away from circles.  Our numeric simulations will corroborate this,
as rectangle-like domains do indeed dominate in this regime.  Finally, it is
important to note previous calculations of the stability
of isolated stripes \cite{CM80a,EDCFB98,KJM95}.  In our terms, they found that the critical
width of a stripe, that is, the largest width for which an infinite stripe
becomes unstable, is given by $d_y=e^{\Lambda+\gamma+2}$, where $\gamma$ is
Euler's constant.  For an energy--minimized rectangle in the high aspect-ratio
limit, its width is given by $w_\rec=d_y=e^{\Lambda+1}$, which is strictly
less than the critical value for all values of $\Lambda$.  This suggests that
the stripe-like portions of branching structures observed in these systems are
stable to perturbation.

Some researchers studying thin ferromagnetic layers use a different
set of parameters than those used here \cite{CM80a,JGC94}.  For a magnetic fluid layer with
thickness $h$, surface tension $\sigma$, and magnetic charge density
$M$,  our parameters translate as $\mu=Mh$, $\lambda=\sigma h$, and 
$\Delta\sim e^{-3/2}h$ in the limit of small thickness $h$.  The generalized line tension $\Lambda$ is given, in terms
of these parameters, as
\[\Lambda=\frac32+\frac\sigma{M^2h}-\log\frac
	Rh=\frac32+\frac2{N_B}-\log\frac Rh
\]
where $N_B\equiv M^2h/\sigma$ is the magnetic bond number \cite{JGC94}.

\section{Numerics}

In order to perform numeric simulations of dipole-mediated domains, we
discretized the boundary $\bound\Omega$ in the energy expression
\eqref{energy.final}.  Consider a set of $N$ points $\vec x_i=(x_i,y_i)$, each
equidistant to its adjacent neighbors.  Adjacent points are kept equidistant
to ensure that they are uniformly spread along the domain's perimeter.  This
equidistance condition can be expressed by the $N$ consistency equations
\[
	\frac LN=\norm{\vec x_{i+1}-\vec x_i}.
	\label{equidistant}
\]
Define $\vec\rr_i\equiv\frac12(\vec x_{i+1}+\vec x_i)$ and $\vec
t_i\equiv\frac12(\vec x_{i+1}-\vec x_i)$ to approximate the midpoint and
tangent vectors of the polygonal sides.  The normal vector $\vec n_i$ is
defined to be the outward facing vector orthogonal to $\vec t_i$ and of the
same length.  The simplest discretization of the energy integration given this
boundary discretization is
\[
	F=\Lambda L-\frac12\sum_{i=1}^N\sum_{j=-N/2}^{\frac{N}2-1}\left(
	\frac{\unit t_{i+j}\cdot\unit t_j}{\norm{\vec\rr_{i+j}-\vec\rr_j}}-\frac1{\frac LN\abs
	j}\right)\frac{L^2}{N^2}-L\log L+\sum_{i=1}^N\vec\Pi(\vec
	x_i)\cdot\unit n_i\frac LN,
\]
where the hats on some vectors denote that they have been normalized to unit
length.  Using \eqref{equidistant}, for instance, one can see that $\vec
t_i=L/N\;\unit t_i$.
The expression above can be simplified considerably by computing the sum over
the second term in the summand, yielding
\[
	F=\left (\Lambda+H_{\frac{N}2 -1} +\frac1N \right )L-\frac12\sum_{i=1}^N\sum_{\substack{j=1\\j\neq
	i}}^N\frac{\vec t_i\cdot\vec t_j}{\norm{\vec\rr_i-\vec\rr_j}}-L\log L+\sum_{i=1}^N\vec\Pi(\vec
	x_i)\cdot\vec n_i.
	\label{disc.energy}
\]
where $H_m=\sum_{j=1}^m \frac{1}{j}$ is the $m$th harmonic number.  In order to ensure that the area of
a domain stays constant, the boundary points must fulfill the consistency
expression
\[
	\pi=\frac12\norm{\sum_{i=1}^N\vec x_{i+1}\times\vec
	x_i}=\frac12\Abs{\sum_{i=1}^N(x_iy_{i+1}-x_{i+1}y_i)}.
	\label{area}
\]
We are now looking at a problem of constrained optimization; we use Lagrange
multipliers to minimize \eqref{disc.energy} under the constraints
\eqref{equidistant} and \eqref{area}.  The Lagrangian for such a constrained
system is given by
\[
	\Ell=F
	-\lambda_0\left[\pi-\frac12\Abs{\sum_{i=1}^N(x_iy_{i+1}-x_{i+1}y_i)}\right]
	-\sum_{i=1}^N\lambda_i\left[\frac{L^2}{N^2}-\norm{\vec x_{i+1}-\vec
	x_i}^2\right],
\]
where $\lambda_0,\ldots,\lambda_N$ are the Lagrange multipliers.  We minimize
the energy of this discrete system to investigate stable domain configurations
using a modified version of the Levenberg–Marquardt algorithm (LMA).  Normally
the LMA corresponds to a modified Newton's method where a multiple of the
identity matrix is added to the Hessian before solving for the step size.
When this multiple is very large, the algorithm acts like gradient following,
minimizing energy as opposed to Newton's method which converges to any
critical point.  However, in a system containing Lagrange multipliers as
variables, minimization of the energy with respect to all variables is
impossible, since the multipliers can increase without bound and drive the
algorithm to diverge.  Therefore, we use LMA where, instead of an identity
matrix, we add a block identity matrix to the Hessian, so that the Lagrangian
is minimized with respect to the physical variables while respecting the
constraints associated with the Lagrange multipliers.  If $\vec
z=[x_1,\ldots,x_N,y_1,\ldots,y_N,\ell]$ is the vector of physical variables
and $\vec\lambda=[\lambda_0,\ldots,\lambda_N]$ is that of Lagrange
multipliers, the system is described by the state vector $[\vec
z,\vec\lambda]$.  In our modified algorithm, a step is given by
\[
	\begin{bmatrix}
		\Delta\vec z\\\Delta\vec\lambda
	\end{bmatrix}
	=
	\alpha\left(
	H_\Ell+
	\eta\begin{bmatrix}
		I_{2N+1}&0\\0&0
	\end{bmatrix}
	\right)^{-1}
	\del\Ell
\]
where $H_\Ell$ and $\del\Ell$ are the Hessian and gradient of the Lagrangian
at the previous state, $I_n$ is the $n\times n$ identity matrix, $\alpha$ is
chosen using the Armijo rule \cite{B99}.  The parameter $\eta$ is set to some
initial value $\eta_0$, and then is decremented as the gradient of $\Ell$ dips
below some pre-selected value.

\begin{figure}
	\centering
	\hspace{22pt}\includegraphics{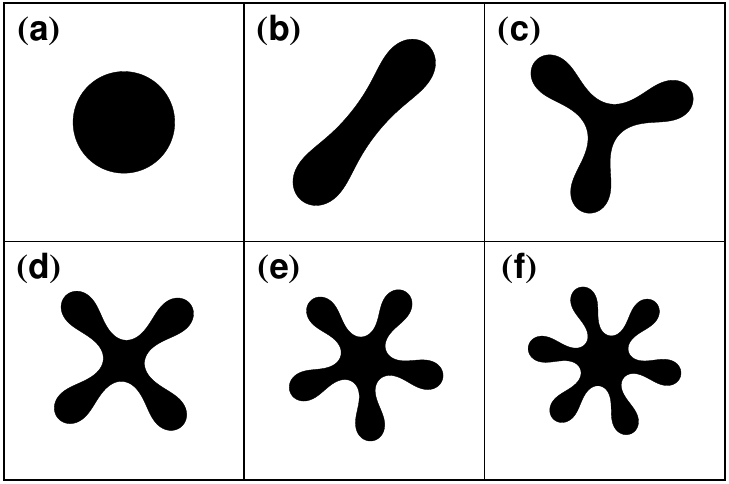}
	\includegraphics{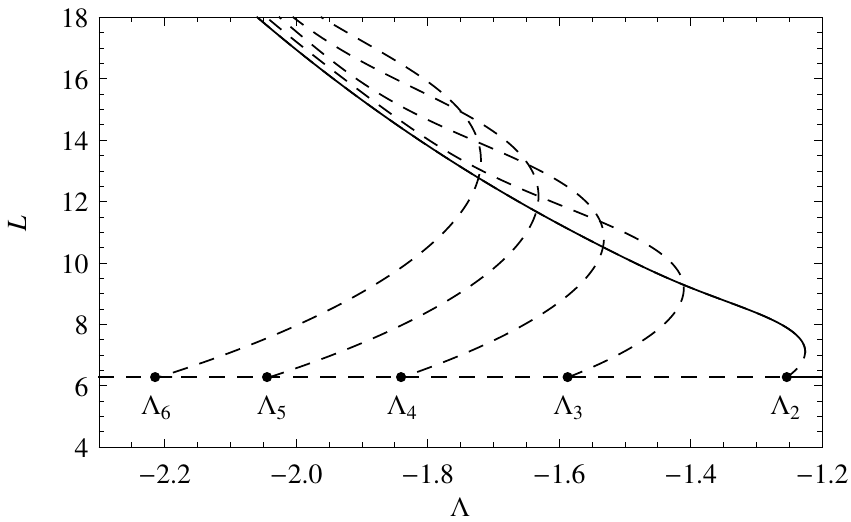}
	\caption{Bottom: The perimeters of the first five harmonic bifurcations from
		a circular domain.  The black dots represent the theoretical bifurcation
		points $\Lambda_n$, the solid lines denote stable numeric solutions, and
		the dashed lines denote unstable numeric solutions. Top: A circular domain
		alongside those bifurcations.  These shapes were taken with $\Lambda$
		values of (a) $-1.2$, (b) $-1.38$, (c) $-1.52$, (d) $-1.65$, (e) $-1.69$,
		and (f) $-1.77$.
	}
	\label{fig:harmonic.bifur}
\end{figure}

\begin{figure}
	\centering
	\includegraphics{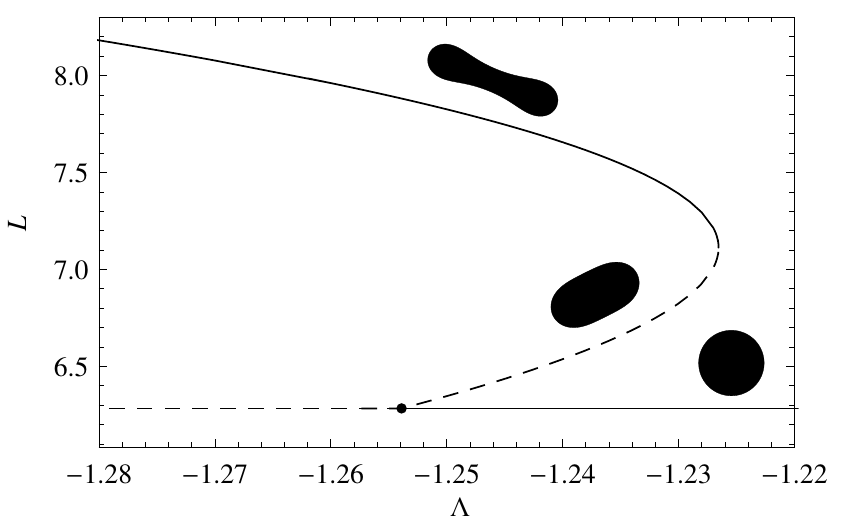}
	\caption{The subcritical bifurcation of the stripe/dogbone from the circle.  The solid
		lines denote stable numeric solutions and the dashed lines denote unstable
		ones. The subcritical branch of solutions regains stability in a fold bifurcation at $\Lambda\simeq-1.227$. }
	\label{fig:dogbone}
\end{figure}

In the absence of an external potential ($\vec\Pi(r)=0$), we used continuation
in $\Lambda$ to examine the domain shapes which are stable, i.e., energy
minimizers.  For sufficiently large $\Lambda$ the circular domain is the
unambiguous global minimizer.  Once $\Lambda\simeq\Lambda_2$, the phase space
becomes far more interesting.  We were able to follow the harmonic
bifurcations from a circular domain onto their solution branches.  The first
five harmonic bifurcations can be seen in Fig.~\ref{fig:harmonic.bifur}(a--f),
and their branches as represented by the perimeter $L$ are plotted in the same
figure.  Notice that all harmonic bifurcations exhibit the same subcritical
branching behavior.  Stability is recovered for the $\Lambda_2$ branch, which
corresponds to the circle to dogbone transition (see Fig.~\ref{fig:dogbone}).
This subcritical behavior has been observed previously by Cebers and his
collaborators and is responsible for the hysteresis in dogbone
formation and relaxation previously observed by Jackson and his collaborators
\cite{CZ83,CD99,JH07,J08}.  We find that the value of $\Lambda$ at the tip of the upper
branch is $\Lambda\simeq-1.227$.  Numerically, the circle appears to be the
global attractor above this point.  Harmonic branches of fourfold and higher
symmetries have been shown by Cebers et al.\ to decay into lower-symmetry
shapes in magnetohydrodynamic simulations as a result of so-called vertex
splitting instability \cite{CD95,CD99}.  Our eigenvalue analysis confirms that
these domains are unstable.  We also found that the branch of threefold
symmetry is unstable, despite the fact that it has not decayed in previous
magnetohydrodynamic simulations \cite{CZ83,CD99}.  This inconsistency is due
to the fact that the decay mechanism for the threefold symmetry is not vertex
splitting---which is geometrically forbidden---and is instead caused by the
atrophy of one or two of the arms, which (as we will discuss below) is a much weaker instability and would
manifest itself on a much larger timescale than the vertex splitting in
dynamic simulation. Moreover, previous simulations use different strategies for regularizing the dipole energy; it is plausible that this these changes effect the observed stability for, say, a small but finite magnetic layer thickness.

\begin{figure}
	\centering
	\includegraphics{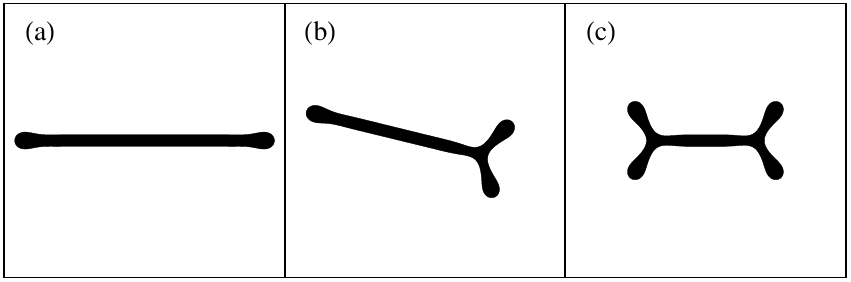}
	\caption{Representatives of (a) stripe, (b) forked, and (c) doubly forked
		domain morphologies at $\Lambda=-2$.  These appear to be the only stable
		morphologies in the range of $\Lambda$ we have investigated in the absence of a random energy background.}
	\label{fig:minimizers}
\end{figure}

\begin{figure}
	\centering
	\includegraphics{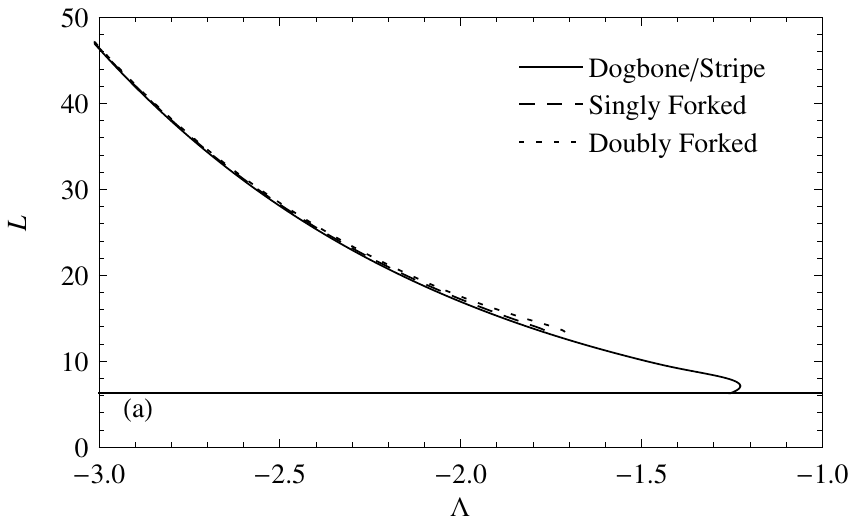}\\
	\vspace{-19pt}
	\includegraphics{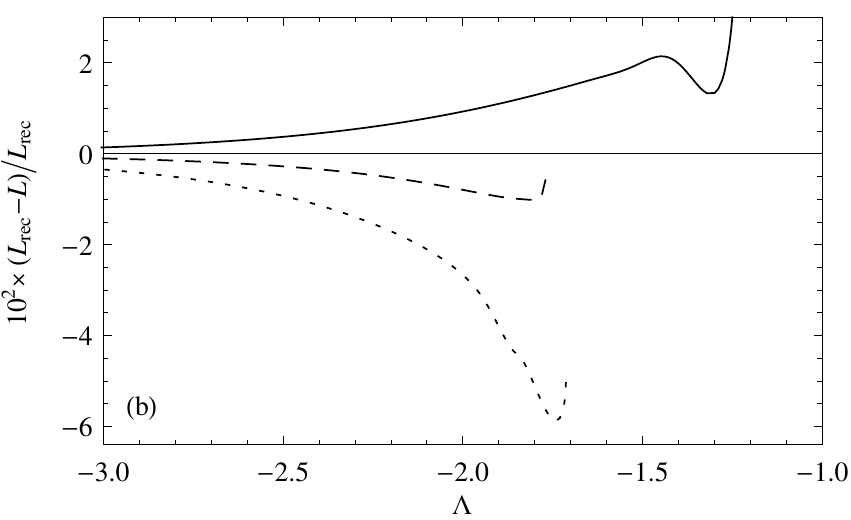}
	\caption{The asymptotic behavior of the perimeter of the three stable domain
		morphologies for $N=8196$.  (a) The perimeter of each morphology as a
		function of $\Lambda$. (b) The relative error between the perimeter of
		each morphology and $L_\rec$, the asymptotic rectangle perimeter.}
	\label{fig:stable_asymptotics}
\end{figure}

If the dogbone is allowed to adiabatically evolve with decreasing $\Lambda$,
it becomes long and stripe-like, very much like the rectangle we considered in
the analytic section.  In particular, the stripe is stable, and we suspect it
is the global minimizer in the regime where the circle is no longer stable.
Through bifurcation following on higher harmonic branches, we found two other
stable morphologies: the forked and doubly forked domains.  These are
represented in Fig.~\ref{fig:minimizers}.  Notice that all three of these
solutions appear like rectangles with various modifications to their ends.  In
fact, all three stable morphologies evolve in a similar way, becoming very
long and stripe-like with large $-\Lambda$.  The perimeter of these domains as
a function of $\Lambda$ can be seen in Fig.~\ref{fig:stable_asymptotics}(a).
The perimeters of all three increase exponentially, and in fact almost
identically to the analytic rectangle perimeter $L_\rec(\Lambda)$.  The close
connection between the perimeters of these stable shapes and that of
$L_\rec(\Lambda)$ can be seen in Fig.~\ref{fig:stable_asymptotics}(b), which
shows the relative error between the perimeters of each stable shape and
$L_\rec(\Lambda)$.  As can be seen from that figure, the difference between
the perimeters of these shapes and the analytic rectangle becomes less than
2\% for $\Lambda\simeq-2$ and less than 1\% at $\Lambda\simeq-2.5$.  In fact,
even the unstable higher harmonic bifurcations behave like this, approaching
asymptotically the rectangle perimeter as $\Lambda$ becomes more negative.

\begin{figure}
	\centering
	\includegraphics{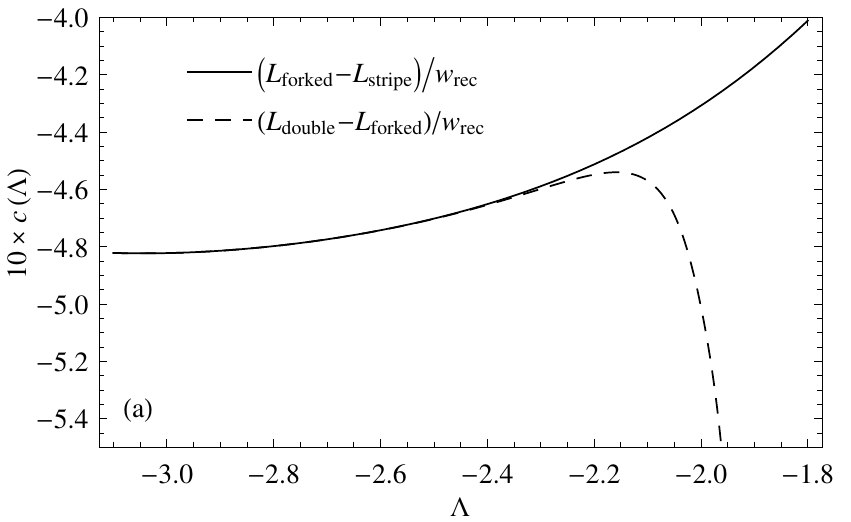}\\
	\vspace{-19pt}
	\includegraphics{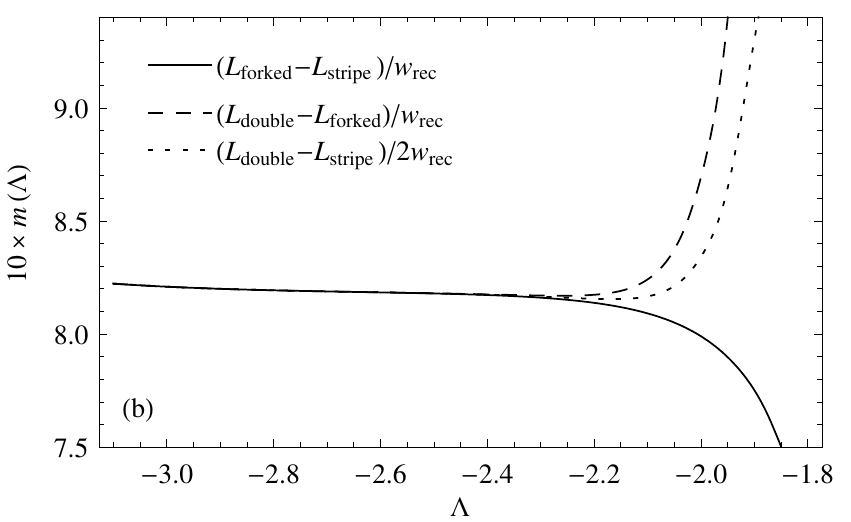}
	\caption{The ratios of the rectangle width $w_\rec$ to linear combinations of 
		the stable perimeters that correspond, in the limit of negative $\Lambda$,
		to the constants (a) $c$ and (b) $m$.
	}
	\label{fig:const_slope}
\end{figure}

Note further that the stripe has a slightly lower perimeter than the
rectangle, while the forked and doubly forked domains have progressively
higher perimeters.  The central bulk of the stripe is geometrically identical
to the rectangle in all respects.  Therefore, the curved ends of the stripe
domain must be responsible for the deviation.  These ends have a size
proportional to the width of the stripe, which is in turn proportional to
$w_\rec(\Lambda)$, the asymptotic rectangle width.  The difference between the
perimeters of the stripe domain and the rectangle should likewise be
proportional to the size of the anomalous ends.  Hence, in the limit of large
negative $\Lambda$, the expressions
\begin{align}
	\frac{L_\stripe-L_\rec}{w_\rec}&&
	\frac{L_\double-2L_\forked-L_\rec}{w_\rec}
	\label{c.values}
\end{align}
should go to the same constant $c$, loosely the energy cost per endcap. This
is a nontrivial statement, since $w_\rec$ decreases exponentially as $\Lambda$
becomes more negative, so $L_\stripe-L_\rec$ will have to decrease equally
exponentially in order for $c$ to converge.  However, this is exactly what we
see.  Both expressions in \eqref{c.values}  can be seen plotted as a function
of $\Lambda$ in Fig.~\ref{fig:const_slope}(a).  The constant itself can be
roughly determined by sampling along the relatively constant region between
$-2.8$ and $-3.1$ and averaging, yielding $c=-0.482\pm0.001$.

In addition, we need to account for the perimeter differences of the forked
and doubly forked domains.  When a junction is added to a stripe-like shape,
another anomalous end is added.  Like the ends, the size of the junction
itself also scales with the width of the domain.  Therefore, we should expect
that there is a \emph{cost per threefold junction} which scales like
$w_\rec(\Lambda)$, so that in the limit of large negative $\Lambda$, the
expressions
\begin{align}
	\frac{L_\forked-L_\stripe}{w_\rec}&&
	\frac{L_\double-L_\forked}{w_\rec}&&
	\frac{L_\double-L_\stripe}{2w_\rec}
\end{align}
should go to the same constant $m$.  As can be seen in
Fig.~\ref{fig:const_slope}(b), this is indeed the case.  All three ratios
tend to the same constant, which can be determined to be $m=0.819\pm0.001$
(see \cite{K14} for details).

Given this description, one might imagine that the perimeter of any simply
connected domain with $n$ threefold junctions (and no junctions of higher
order) will be, 
\[
	L\simeq L_\rec+(c+mn)w_\rec
	\label{model.ini}
\]
for $\Lambda$ sufficiently negative.
This is a remarkably simple characterization of complicated domain
structure, but, as we will see, it indeed holds for domains which resemble
the intricate structure of those seen in experiment.  Though this model
necessarily restricts itself to domains with threefold junctions, recall that
we only found stable shapes with threefold junctions.  As it turns out,
junctions of higher order are never seen in stable shapes in our
numerics and rarely seen in experimental domains due to the vertex splitting
instability, which highly disfavors vertex symmetries of fourfold symmetry and
higher \cite{CD95,CD99}.

\begin{figure}
	\centering
	\includegraphics[width=2in]{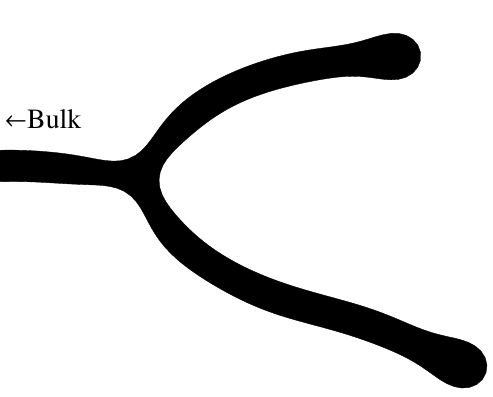}
	\caption{An example of two branches in a more complex branching domain.}
	\label{fig:branching}
\end{figure}

Unfortunately, the stable domain structures seen in Fig.~\ref{fig:minimizers}
lack many of the qualitative properties seen in experiment, e.g., branching
structure, asymmetry, and snaking behavior.  We suspect that this is because,
in experimental settings, there is a nonzero effective background potential
$V(\vec r)$.  This could come from small inhomogeneities of the substrate,
fringing or imperfect applied fields, etc.  These imperfections can pin nearly
stable domains.   We can estimate the size of the potential needed for this
pinning by considering two branches of a typical structure, like those in
Fig.~\ref{fig:branching}.  Our analysis of the threefold harmonic shape
suggests such a configuration is unstable and will decay by shortening one
branch down into the other.  We wish to find the energy gradient associated
with this decay.  Consider a small cross section of the upper branch and
compute the energy it takes to move this piece onto the lower branch.  Since
such a move conserves the perimeter of the shape, the line tension and
logarithmic terms in the energy do not change.  The dipole energy of the small
section with respect to the bulk scales like the area of the section,
$w_\rec\cdot\Delta x$, over the cube of the mean distance of that section from
the rest of material, which we expect to scale like $L_\rec$.  There is a
scaling constant $c_1$ that depends on the geometry of the bulk relative to
the upper branch.  Upon moving to the lower branch, the scaling behavior is
identical, but the bulk relation constant changes to $c_2$.  Therefore, we
have
\[\Delta F=\frac{w_\rec\Delta x}{L_\rec^3}(c_2-c_1)\]
Using the known scaling behavior of $L_\rec$ and $w_\rec$, this can be
written
\[\frac{\Delta F}{\Delta x}\sim e^{4(\Lambda+1)}\]
Thus, as $\Lambda$ becomes negative, like it does in the regime where we see
branching structures emerge, the energy gradient which destroys branching
structures becomes smaller exponentially.  In this regime, we should expect to
see branching structures begin to emerge over random backgrounds of even
modest amplitude.

The form of our random energy background is as follows.  First, we choose
positive real numbers $k_0$ and $a_0$ to characterize the scale of the noise
and an integer $M$ to give the number of modes included.  Then, we create a
set of vectors $\{\vec k_i\}$ and sets of scalars $\{a_i\}$ and $\{\phi_i\}$,
where $i=1,\ldots M$.  The $\vec k_i$ are taken from a uniform distribution in
the circle of radius $k_0$ centered at the origin, the $a_i$ are taken
uniformly from the interval $[0,2a_0/M]$, and the $\phi_i$ are taken uniformly
from the interval $[0,2\pi]$.  The background energy is then given by the
density
\[
	V(\vec\rr)=\sum_{i=1}^Ma_i\cos(\vec k_i\cdot\vec\rr+\phi_i).
\]
Consider the function $\vec\Pi:\reals^2\to\reals^2$ defined by
\[
	\vec\Pi(\vec\rr)=\frac12\sum_{i=1}^Ma_i\sin(\vec k_i\cdot\vec\rr+\phi_i)\left[\frac1{k_{ix}},\;\frac1{k_{iy}}\right].
\]
It follows that $\del\cdot\vec\Pi=V$.
This is precisely the condition we have on the external line potential
$\vec\Pi$.  Therefore, the numerical approximation to the energy is given by
\[
	F_{\text{rand}}=\oint_{\partial\Omega}\vec\Pi\cdot\unit
n\,\dd s=\oint_{\partial\Omega}\dd s\;(\Pi_x\hat n_x+\Pi_y\hat
n_y)=\frac12\oint_{\partial\Omega}\sum_{i=1}^Ma_i\sin(\vec k_i\cdot\vec\rr+\phi_i)\left(\frac{\hat t_y}{k_{ix}}-\frac{\hat t_x}{k_{iy}}\right)\dd s,
\]
where we have used $\hat n_x=\hat t_y$ and $\hat n_y=-\hat t_x$, true for the
tangents and normals of positively oriented domains.  This means that, given
the discretization of the domain boundary we used before,
\[F_{\text{rand}}=\frac12\sum_{j=1}^N\sum_{i=1}^Ma_i\sin(\vec k_i\cdot\vec\rr_j+\phi_i)
\left(\frac{y_{j+1}-x_{j}}{k_{ix}}-\frac{x_{j+1}-x_{j}}{k_{iy}}\right),\]
where the first sum is over the points making up the sides of the domain and
the indices are defined cyclically.  We can now simulate domains over such
backgrounds in precisely the same way as we did in the case without the
background.

\begin{figure}
	\centering
	\includegraphics{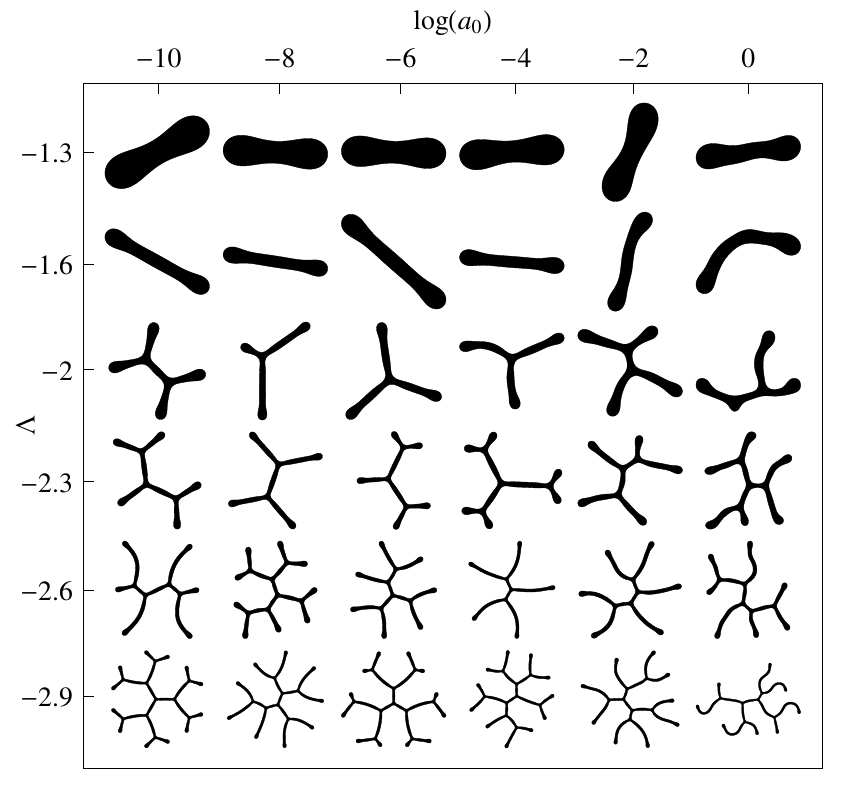}
	\caption{A sampling of stable solutions to our numeric model over random
		external potentials.  Travelling down the vertical axis corresponds to
		decreasing $\Lambda$ and travelling to the right on the horizontal axis corresponds to increasing
		background intensity.}
	\label{fig:menagerie}
\end{figure}

\begin{figure}
	\centering
	\includegraphics{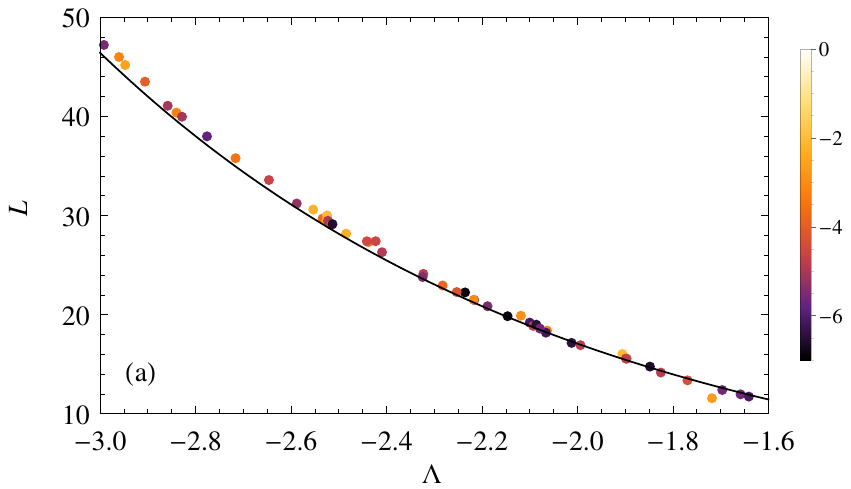}\\
	\vspace{-19pt}
	\includegraphics{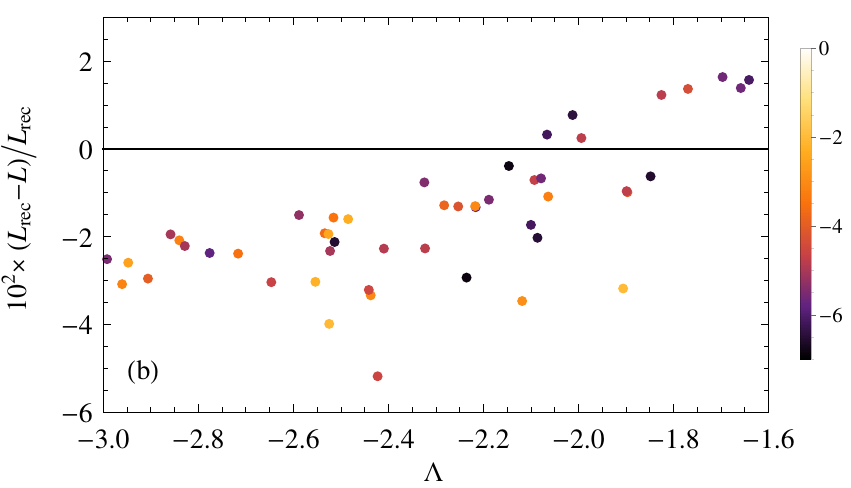}
	\caption{(Color online) The top plot shows the perimeter $L$ of stable
		domains as a function of $\Lambda$.  The color of each point denotes the
		value of $\log_{10}(a_0)$, the order of magnitude of the random
		background, as detailed by the legend.  The solid black line is a plot of
		$L_\rec(\Lambda)$.  The lower plot shows the relative error of each
		perimeter from $L_\rec$.}
	\label{fig:scatter}
\end{figure}

With this modification, we are able to recover many of the qualitative
features seen in experiment that were not seen for the minimizers we observe in the absence of noise.  See, for instance,
Fig.~\ref{fig:menagerie}, which shows samples of these shapes at a variety of
$\Lambda$ values and background intensities.  Moreover, we found that the
perimeters of these shapes continue to correspond, to a large degree, with the
rectangle relationship $L_\rec(\Lambda)$ that we found before.  In
Fig.~\ref{fig:scatter}, we have plotted the perimeter of arbitrary shapes over
a random background as a function of $\Lambda$.  The color of each point
corresponds to the intensity of the random background it was generated in.
Notice first that the nature of the background does not seem to influence
domain perimeter in a regular way.  Next, notice that despite the relative
complexity of these shapes, their perimeters remain very close to the
rectangular idealization.  However, there is an upward trend with increasing
negative $\Lambda$.

\begin{figure}
	\centering
	\includegraphics{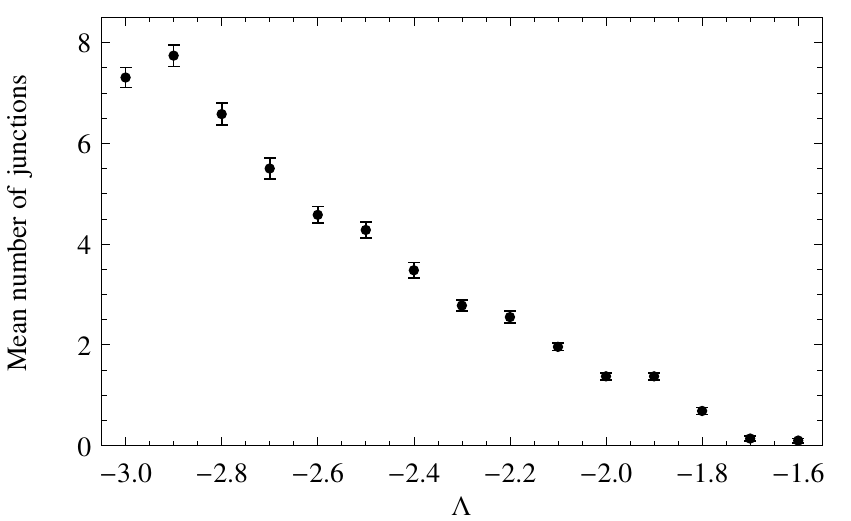}
	\caption{The average number of junctions in numeric domains as a
		function $\Lambda$.  The error bars denote standard error.}
	\label{fig:junctions}
\end{figure}

\begin{figure}
	\centering
	\includegraphics{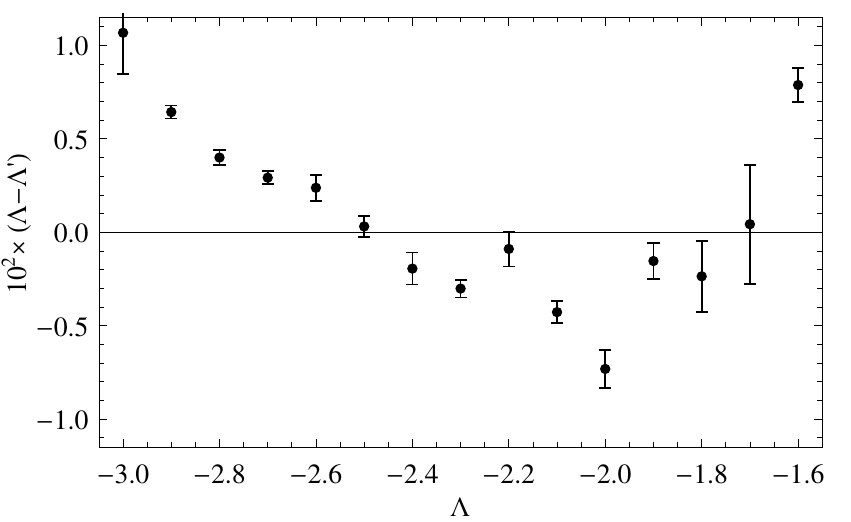}
	\caption{The difference between the generating value $\Lambda$ and the mean
		predicted value $\Lambda'$ for sets of $50$ domains.  The error bars denote
		standard error.}
	\label{fig:rule.1200}
\end{figure}

A similar upward trend exists in another shape-relevant morphological
parameter: the number of junctions in the domain.  This trend is shown in
Fig.~\ref{fig:junctions}.  Given the simple model \eqref{model.ini}, one would
expect a greater number of junctions to correspondingly cause inflation in the
observed perimeter from that of the rectangle.  By inverting that model, we
can make a prediction $\Lambda'$ of a domain's true value of $\Lambda$, i.e.,
that at which it was generated.  This prediction is given by
\[\Lambda'=\log\left[\frac{L-\sqrt{L^2-8\pi(c+mn)}}{2(c+mn)}\right]-1.
\label{model}
\]
We tested this model at values of $\Lambda$ between $-3$ and $-1.6$ for sets
of $50$ domains minimized over random backgrounds.  Fig.~\ref{fig:rule.1200}
shows the error in those predictions.  As can be seen there, for all $\Lambda$
tested the error in our model was less than 1\%.  The upward trend for
$\Lambda\sim-3$ may be due to numerical under-resolution \cite{K14}.

\section{Conclusions}

In this work, we have developed a way to express the energy of a
dipole-mediated system that depends only on a single non-dimensional
parameter, $\Lambda$.  Numeric simulations using energy minimization were
developed.  We used these simulations to track the bifurcations of domains
from a circle, and resolve subcritical branches for the first five harmonic
bifurcations, and in particular for that of the circle to dogbone transition.
Using the same methodology, we found three stable domain morphologies beyond
the circle, all of which resemble a rectangular domain in appearance and
behavior.  Among these, the stripe, which evolves from a dogbone, is suspected
to be the global energy minimizer in the unstable-circle regime. 

The fact that these observed domains lack the qualitative features of
experimental domains led to the conclusion that those features necessarily
depend on the presence of an imperfect background energy landscape, and we
confirmed this by recovering those features in our numerics.  Using these
domains, we found that a simple model suggested by the stable domains
continues to work well in deriving the value of $\Lambda$ from the shape of an
arbitrary domain.  This model is especially powerful, because it only relies
on the area-normalized perimeter and number of junctions present in the shape.
These features can extracted from photographs of experiments, and so recovery
of $\Lambda$, which contains ratios of physical variables, is straightforward
in practice.  An experimentalist could use this technique while varying some
known parameter, e.g., the magnetic field or domain area, to work out other,
unknown parameters by a fit of the system's $\Lambda$-dependence.  We hope to
explore the possibility of our model being used in this way through
collaboration with experimentalists.

\section{Acknowledgements}

This work was done in part under NSF grants CBET-07306630 and DMS-1009633.  The authors would
like to thank Elizabeth Mann and David Jackson for the
experimental images they generously contributed, the Mathematics and Biology
departments of Harvey Mudd College for the use of their computing resources, and
Professor Chad Higdon-Topaz for his support.

\bibliography{article}

\begin{thebibliography}{33}%
\makeatletter
\providecommand \@ifxundefined [1]{%
 \@ifx{#1\undefined}
}%
\providecommand \@ifnum [1]{%
 \ifnum #1\expandafter \@firstoftwo
 \else \expandafter \@secondoftwo
 \fi
}%
\providecommand \@ifx [1]{%
 \ifx #1\expandafter \@firstoftwo
 \else \expandafter \@secondoftwo
 \fi
}%
\providecommand \natexlab [1]{#1}%
\providecommand \enquote  [1]{``#1''}%
\providecommand \bibnamefont  [1]{#1}%
\providecommand \bibfnamefont [1]{#1}%
\providecommand \citenamefont [1]{#1}%
\providecommand \href@noop [0]{\@secondoftwo}%
\providecommand \href [0]{\begingroup \@sanitize@url \@href}%
\providecommand \@href[1]{\@@startlink{#1}\@@href}%
\providecommand \@@href[1]{\endgroup#1\@@endlink}%
\providecommand \@sanitize@url [0]{\catcode `\\12\catcode `\$12\catcode
  `\&12\catcode `\#12\catcode `\^12\catcode `\_12\catcode `\%12\relax}%
\providecommand \@@startlink[1]{}%
\providecommand \@@endlink[0]{}%
\providecommand \url  [0]{\begingroup\@sanitize@url \@url }%
\providecommand \@url [1]{\endgroup\@href {#1}{\urlprefix }}%
\providecommand \urlprefix  [0]{URL }%
\providecommand \Eprint [0]{\href }%
\providecommand \doibase [0]{http://dx.doi.org/}%
\providecommand \selectlanguage [0]{\@gobble}%
\providecommand \bibinfo  [0]{\@secondoftwo}%
\providecommand \bibfield  [0]{\@secondoftwo}%
\providecommand \translation [1]{[#1]}%
\providecommand \BibitemOpen [0]{}%
\providecommand \bibitemStop [0]{}%
\providecommand \bibitemNoStop [0]{.\EOS\space}%
\providecommand \EOS [0]{\spacefactor3000\relax}%
\providecommand \BibitemShut  [1]{\csname bibitem#1\endcsname}%
\let\auto@bib@innerbib\@empty
\bibitem [{\citenamefont {Cebers}(1980)}]{C80}%
  \BibitemOpen
  \bibfield  {author} {\bibinfo {author} {\bibfnamefont {A.}~\bibnamefont
  {Cebers}},\ }\href@noop {} {\bibfield  {journal} {\bibinfo  {journal}
  {Magnetohydrodynamics}\ }\textbf {\bibinfo {volume} {16}},\ \bibinfo {pages}
  {236} (\bibinfo {year} {1980})}\BibitemShut {NoStop}%
\bibitem [{\citenamefont {Cebers}\ and\ \citenamefont
  {Maiorov}(1980{\natexlab{a}})}]{CM80a}%
  \BibitemOpen
  \bibfield  {author} {\bibinfo {author} {\bibfnamefont {A.}~\bibnamefont
  {Cebers}}\ and\ \bibinfo {author} {\bibfnamefont {M.}~\bibnamefont
  {Maiorov}},\ }\href@noop {} {\bibfield  {journal} {\bibinfo  {journal}
  {Magnetohydrodynamics}\ }\textbf {\bibinfo {volume} {16}},\ \bibinfo {pages}
  {21} (\bibinfo {year} {1980}{\natexlab{a}})}\BibitemShut {NoStop}%
\bibitem [{\citenamefont {Cebers}\ and\ \citenamefont {Mayorov}(1980)}]{CM80b}%
  \BibitemOpen
  \bibfield  {author} {\bibinfo {author} {\bibfnamefont {A.}~\bibnamefont
  {Cebers}}\ and\ \bibinfo {author} {\bibfnamefont {M.}~\bibnamefont
  {Mayorov}},\ }\href@noop {} {\bibfield  {journal} {\bibinfo  {journal}
  {Magnetohydrodynamics}\ }\textbf {\bibinfo {volume} {16}},\ \bibinfo {pages}
  {231} (\bibinfo {year} {1980})}\BibitemShut {NoStop}%
\bibitem [{\citenamefont {Cebers}\ and\ \citenamefont
  {Maiorov}(1980{\natexlab{b}})}]{CM80c}%
  \BibitemOpen
  \bibfield  {author} {\bibinfo {author} {\bibfnamefont {A.}~\bibnamefont
  {Cebers}}\ and\ \bibinfo {author} {\bibfnamefont {M.}~\bibnamefont
  {Maiorov}},\ }\href@noop {} {\bibfield  {journal} {\bibinfo  {journal}
  {Magnetohydrodynamics}\ }\textbf {\bibinfo {volume} {16}},\ \bibinfo {pages}
  {126} (\bibinfo {year} {1980}{\natexlab{b}})}\BibitemShut {NoStop}%
\bibitem [{\citenamefont {Cebers}\ and\ \citenamefont {Zemitis}(1983)}]{CZ83}%
  \BibitemOpen
  \bibfield  {author} {\bibinfo {author} {\bibfnamefont {A.}~\bibnamefont
  {Cebers}}\ and\ \bibinfo {author} {\bibfnamefont {A.}~\bibnamefont
  {Zemitis}},\ }\href@noop {} {\bibfield  {journal} {\bibinfo  {journal}
  {Magnetohydrodynamics}\ }\textbf {\bibinfo {volume} {19}},\ \bibinfo {pages}
  {360} (\bibinfo {year} {1983})}\BibitemShut {NoStop}%
\bibitem [{\citenamefont {McConnell}\ and\ \citenamefont {Moy}(1988)}]{MM88}%
  \BibitemOpen
  \bibfield  {author} {\bibinfo {author} {\bibfnamefont {H.~M.}\ \bibnamefont
  {McConnell}}\ and\ \bibinfo {author} {\bibfnamefont {V.~T.}\ \bibnamefont
  {Moy}},\ }\href {\doibase 10.1021/j100326a053} {\bibfield  {journal}
  {\bibinfo  {journal} {The Journal of Physical Chemistry}\ }\textbf {\bibinfo
  {volume} {92}},\ \bibinfo {pages} {4520} (\bibinfo {year} {1988})},\ \Eprint
  {http://arxiv.org/abs/http://pubs.acs.org/doi/pdf/10.1021/j100326a053}
  {http://pubs.acs.org/doi/pdf/10.1021/j100326a053} \BibitemShut {NoStop}%
\bibitem [{\citenamefont {Vanderlick}\ and\ \citenamefont
  {Moehwald}(1990)}]{VM90}%
  \BibitemOpen
  \bibfield  {author} {\bibinfo {author} {\bibfnamefont {T.}~\bibnamefont
  {Vanderlick}}\ and\ \bibinfo {author} {\bibfnamefont {H.}~\bibnamefont
  {Moehwald}},\ }\href@noop {} {\bibfield  {journal} {\bibinfo  {journal}
  {Journal of Physical Chemistry}\ }\textbf {\bibinfo {volume} {94}},\ \bibinfo
  {pages} {886} (\bibinfo {year} {1990})}\BibitemShut {NoStop}%
\bibitem [{\citenamefont {McConnell}(1990)}]{M90}%
  \BibitemOpen
  \bibfield  {author} {\bibinfo {author} {\bibfnamefont {H.~M.}\ \bibnamefont
  {McConnell}},\ }\href@noop {} {\bibfield  {journal} {\bibinfo  {journal}
  {Journal of Physical Chemistry}\ }\textbf {\bibinfo {volume} {94}},\ \bibinfo
  {pages} {4728} (\bibinfo {year} {1990})}\BibitemShut {NoStop}%
\bibitem [{\citenamefont {McConnell}\ and\ \citenamefont
  {de~Koker}(1992)}]{MK92}%
  \BibitemOpen
  \bibfield  {author} {\bibinfo {author} {\bibfnamefont {H.~M.}\ \bibnamefont
  {McConnell}}\ and\ \bibinfo {author} {\bibfnamefont {R.}~\bibnamefont
  {de~Koker}},\ }\href@noop {} {\bibfield  {journal} {\bibinfo  {journal} {The
  Journal of Physical Chemistry}\ }\textbf {\bibinfo {volume} {96}},\ \bibinfo
  {pages} {7101} (\bibinfo {year} {1992})}\BibitemShut {NoStop}%
\bibitem [{\citenamefont {Langer}\ \emph {et~al.}(1992)\citenamefont {Langer},
  \citenamefont {Goldstein},\ and\ \citenamefont {Jackson}}]{LGJ92}%
  \BibitemOpen
  \bibfield  {author} {\bibinfo {author} {\bibfnamefont {S.~A.}\ \bibnamefont
  {Langer}}, \bibinfo {author} {\bibfnamefont {R.~E.}\ \bibnamefont
  {Goldstein}}, \ and\ \bibinfo {author} {\bibfnamefont {D.~P.}\ \bibnamefont
  {Jackson}},\ }\href@noop {} {\bibfield  {journal} {\bibinfo  {journal}
  {Physical Review A}\ }\textbf {\bibinfo {volume} {46}},\ \bibinfo {pages}
  {4894} (\bibinfo {year} {1992})}\BibitemShut {NoStop}%
\bibitem [{\citenamefont {Dickstein}\ \emph {et~al.}(1993)\citenamefont
  {Dickstein}, \citenamefont {Erramilli}, \citenamefont {Goldstein},
  \citenamefont {Jackson},\ and\ \citenamefont {Langer}}]{DEGJL93}%
  \BibitemOpen
  \bibfield  {author} {\bibinfo {author} {\bibfnamefont {A.~J.}\ \bibnamefont
  {Dickstein}}, \bibinfo {author} {\bibfnamefont {S.}~\bibnamefont
  {Erramilli}}, \bibinfo {author} {\bibfnamefont {R.~E.}\ \bibnamefont
  {Goldstein}}, \bibinfo {author} {\bibfnamefont {D.~P.}\ \bibnamefont
  {Jackson}}, \ and\ \bibinfo {author} {\bibfnamefont {S.~A.}\ \bibnamefont
  {Langer}},\ }\href {\doibase 10.1126/science.261.5124.1012} {\bibfield
  {journal} {\bibinfo  {journal} {Science}\ }\textbf {\bibinfo {volume}
  {261}},\ \bibinfo {pages} {1012} (\bibinfo {year} {1993})},\ \Eprint
  {http://arxiv.org/abs/http://www.sciencemag.org/content/261/5124/1012.full.pdf}
  {http://www.sciencemag.org/content/261/5124/1012.full.pdf} \BibitemShut
  {NoStop}%
\bibitem [{\citenamefont {Goldstein}\ and\ \citenamefont
  {Jackson}(1994)}]{GJ94}%
  \BibitemOpen
  \bibfield  {author} {\bibinfo {author} {\bibfnamefont {R.~E.}\ \bibnamefont
  {Goldstein}}\ and\ \bibinfo {author} {\bibfnamefont {D.~P.}\ \bibnamefont
  {Jackson}},\ }\href@noop {} {\bibfield  {journal} {\bibinfo  {journal} {The
  Journal of Physical Chemistry}\ }\textbf {\bibinfo {volume} {98}},\ \bibinfo
  {pages} {9626} (\bibinfo {year} {1994})}\BibitemShut {NoStop}%
\bibitem [{\citenamefont {Jackson}\ \emph {et~al.}(1994)\citenamefont
  {Jackson}, \citenamefont {Goldstein},\ and\ \citenamefont {Cebers}}]{JGC94}%
  \BibitemOpen
  \bibfield  {author} {\bibinfo {author} {\bibfnamefont {D.~P.}\ \bibnamefont
  {Jackson}}, \bibinfo {author} {\bibfnamefont {R.~E.}\ \bibnamefont
  {Goldstein}}, \ and\ \bibinfo {author} {\bibfnamefont {A.~O.}\ \bibnamefont
  {Cebers}},\ }\href@noop {} {\bibfield  {journal} {\bibinfo  {journal}
  {Physical Review E}\ }\textbf {\bibinfo {volume} {50}},\ \bibinfo {pages}
  {298} (\bibinfo {year} {1994})}\BibitemShut {NoStop}%
\bibitem [{\citenamefont {de~Koker}\ and\ \citenamefont
  {McConnell}(1994)}]{KM94}%
  \BibitemOpen
  \bibfield  {author} {\bibinfo {author} {\bibfnamefont {R.}~\bibnamefont
  {de~Koker}}\ and\ \bibinfo {author} {\bibfnamefont {H.~M.}\ \bibnamefont
  {McConnell}},\ }\href@noop {} {\bibfield  {journal} {\bibinfo  {journal} {The
  Journal of Physical Chemistry}\ }\textbf {\bibinfo {volume} {98}},\ \bibinfo
  {pages} {5389} (\bibinfo {year} {1994})}\BibitemShut {NoStop}%
\bibitem [{\citenamefont {de~Koker}\ \emph {et~al.}(1995)\citenamefont
  {de~Koker}, \citenamefont {Jiang},\ and\ \citenamefont {McConnell}}]{KJM95}%
  \BibitemOpen
  \bibfield  {author} {\bibinfo {author} {\bibfnamefont {R.}~\bibnamefont
  {de~Koker}}, \bibinfo {author} {\bibfnamefont {W.}~\bibnamefont {Jiang}}, \
  and\ \bibinfo {author} {\bibfnamefont {H.~M.}\ \bibnamefont {McConnell}},\
  }\href@noop {} {\bibfield  {journal} {\bibinfo  {journal} {The Journal of
  Physical Chemistry}\ }\textbf {\bibinfo {volume} {99}},\ \bibinfo {pages}
  {6251} (\bibinfo {year} {1995})}\BibitemShut {NoStop}%
\bibitem [{\citenamefont {Lubensky}\ and\ \citenamefont
  {Goldstein}(1996)}]{LG95}%
  \BibitemOpen
  \bibfield  {author} {\bibinfo {author} {\bibfnamefont {D.~K.}\ \bibnamefont
  {Lubensky}}\ and\ \bibinfo {author} {\bibfnamefont {R.~E.}\ \bibnamefont
  {Goldstein}},\ }\href@noop {} {\bibfield  {journal} {\bibinfo  {journal}
  {Physics of Fluids (1994-present)}\ }\textbf {\bibinfo {volume} {8}},\
  \bibinfo {pages} {843} (\bibinfo {year} {1996})}\BibitemShut {NoStop}%
\bibitem [{\citenamefont {Seul}\ and\ \citenamefont {Andelman}(1995)}]{SA95}%
  \BibitemOpen
  \bibfield  {author} {\bibinfo {author} {\bibfnamefont {M.}~\bibnamefont
  {Seul}}\ and\ \bibinfo {author} {\bibfnamefont {D.}~\bibnamefont
  {Andelman}},\ }\href@noop {} {\bibfield  {journal} {\bibinfo  {journal}
  {Science}\ }\textbf {\bibinfo {volume} {267}},\ \bibinfo {pages} {476}
  (\bibinfo {year} {1995})}\BibitemShut {NoStop}%
\bibitem [{\citenamefont {McConnell}\ and\ \citenamefont
  {de~Koker}(1996)}]{MK96}%
  \BibitemOpen
  \bibfield  {author} {\bibinfo {author} {\bibfnamefont {H.~M.}\ \bibnamefont
  {McConnell}}\ and\ \bibinfo {author} {\bibfnamefont {R.}~\bibnamefont
  {de~Koker}},\ }\href@noop {} {\bibfield  {journal} {\bibinfo  {journal}
  {Langmuir}\ }\textbf {\bibinfo {volume} {12}},\ \bibinfo {pages} {4897}
  (\bibinfo {year} {1996})}\BibitemShut {NoStop}%
\bibitem [{\citenamefont {Cebers}\ and\ \citenamefont {Drikis}(1996)}]{CD95}%
  \BibitemOpen
  \bibfield  {author} {\bibinfo {author} {\bibfnamefont {A.}~\bibnamefont
  {Cebers}}\ and\ \bibinfo {author} {\bibfnamefont {I.}~\bibnamefont
  {Drikis}},\ }\href@noop {} {\bibfield  {journal} {\bibinfo  {journal}
  {Magnetohydrodynamics}\ }\textbf {\bibinfo {volume} {32}},\ \bibinfo {pages}
  {8} (\bibinfo {year} {1996})}\BibitemShut {NoStop}%
\bibitem [{\citenamefont {Cebers}\ and\ \citenamefont {Drikis}(1999)}]{CD99}%
  \BibitemOpen
  \bibfield  {author} {\bibinfo {author} {\bibfnamefont {A.}~\bibnamefont
  {Cebers}}\ and\ \bibinfo {author} {\bibfnamefont {I.}~\bibnamefont
  {Drikis}},\ }in\ \href@noop {} {\emph {\bibinfo {booktitle} {Free Boundary
  Problems: theory and applications}}},\ Vol.\ \bibinfo {volume} {409},\
  \bibinfo {editor} {edited by\ \bibinfo {editor} {\bibfnamefont
  {I.}~\bibnamefont {Athanasopoulos}}, \bibinfo {editor} {\bibfnamefont
  {J.~F.}\ \bibnamefont {Rodrigues}}, \ and\ \bibinfo {editor} {\bibfnamefont
  {G.}~\bibnamefont {Makrakis}}}\ (\bibinfo  {publisher} {CRC Press},\ \bibinfo
  {year} {1999})\ pp.\ \bibinfo {pages} {14--38}\BibitemShut {NoStop}%
\bibitem [{\citenamefont {Mann}\ and\ \citenamefont {Primak}(1999)}]{MP99}%
  \BibitemOpen
  \bibfield  {author} {\bibinfo {author} {\bibfnamefont {E.~K.}\ \bibnamefont
  {Mann}}\ and\ \bibinfo {author} {\bibfnamefont {S.~V.}\ \bibnamefont
  {Primak}},\ }\href@noop {} {\bibfield  {journal} {\bibinfo  {journal}
  {Physical Review Letters}\ }\textbf {\bibinfo {volume} {83}},\ \bibinfo
  {pages} {5397} (\bibinfo {year} {1999})}\BibitemShut {NoStop}%
\bibitem [{\citenamefont {Khattari}\ and\ \citenamefont
  {Fischer}(2002)}]{KF02}%
  \BibitemOpen
  \bibfield  {author} {\bibinfo {author} {\bibfnamefont {Z.}~\bibnamefont
  {Khattari}}\ and\ \bibinfo {author} {\bibfnamefont {T.~M.}\ \bibnamefont
  {Fischer}},\ }\href@noop {} {\bibfield  {journal} {\bibinfo  {journal} {The
  Journal of Physical Chemistry B}\ }\textbf {\bibinfo {volume} {106}},\
  \bibinfo {pages} {1677} (\bibinfo {year} {2002})}\BibitemShut {NoStop}%
\bibitem [{\citenamefont {Elias}\ \emph {et~al.}(1998)\citenamefont {Elias},
  \citenamefont {Drikis}, \citenamefont {Cebers}, \citenamefont {Flament},\
  and\ \citenamefont {Bacri}}]{EDCFB98}%
  \BibitemOpen
  \bibfield  {author} {\bibinfo {author} {\bibfnamefont {F.}~\bibnamefont
  {Elias}}, \bibinfo {author} {\bibfnamefont {I.}~\bibnamefont {Drikis}},
  \bibinfo {author} {\bibfnamefont {A.}~\bibnamefont {Cebers}}, \bibinfo
  {author} {\bibfnamefont {C.}~\bibnamefont {Flament}}, \ and\ \bibinfo
  {author} {\bibfnamefont {J.-C.}\ \bibnamefont {Bacri}},\ }\href@noop {}
  {\bibfield  {journal} {\bibinfo  {journal} {The European Physical Journal
  B-Condensed Matter and Complex Systems}\ }\textbf {\bibinfo {volume} {3}},\
  \bibinfo {pages} {203} (\bibinfo {year} {1998})}\BibitemShut {NoStop}%
\bibitem [{\citenamefont {Otto}(1998)}]{O98}%
  \BibitemOpen
  \bibfield  {author} {\bibinfo {author} {\bibfnamefont {F.}~\bibnamefont
  {Otto}},\ }\href@noop {} {\bibfield  {journal} {\bibinfo  {journal} {Archive
  for Rational Mechanics and Analysis}\ }\textbf {\bibinfo {volume} {141}},\
  \bibinfo {pages} {63} (\bibinfo {year} {1998})}\BibitemShut {NoStop}%
\bibitem [{\citenamefont {Heinig}\ \emph {et~al.}(2004)\citenamefont {Heinig},
  \citenamefont {Helseth},\ and\ \citenamefont {Fischer}}]{HHF04}%
  \BibitemOpen
  \bibfield  {author} {\bibinfo {author} {\bibfnamefont {P.}~\bibnamefont
  {Heinig}}, \bibinfo {author} {\bibfnamefont {L.~E.}\ \bibnamefont {Helseth}},
  \ and\ \bibinfo {author} {\bibfnamefont {T.~M.}\ \bibnamefont {Fischer}},\
  }\href@noop {} {\bibfield  {journal} {\bibinfo  {journal} {New Journal of
  Physics}\ }\textbf {\bibinfo {volume} {6}},\ \bibinfo {pages} {189} (\bibinfo
  {year} {2004})}\BibitemShut {NoStop}%
\bibitem [{\citenamefont {Hillier}\ and\ \citenamefont {Jackson}(2007)}]{JH07}%
  \BibitemOpen
  \bibfield  {author} {\bibinfo {author} {\bibfnamefont {N.~J.}\ \bibnamefont
  {Hillier}}\ and\ \bibinfo {author} {\bibfnamefont {D.~P.}\ \bibnamefont
  {Jackson}},\ }\href@noop {} {\bibfield  {journal} {\bibinfo  {journal}
  {Physical Review E}\ }\textbf {\bibinfo {volume} {75}},\ \bibinfo {pages}
  {036314} (\bibinfo {year} {2007})}\BibitemShut {NoStop}%
\bibitem [{\citenamefont {Jackson}(2008)}]{J08}%
  \BibitemOpen
  \bibfield  {author} {\bibinfo {author} {\bibfnamefont {D.~P.}\ \bibnamefont
  {Jackson}},\ }\href@noop {} {\bibfield  {journal} {\bibinfo  {journal}
  {Journal of Physics: Condensed Matter}\ }\textbf {\bibinfo {volume} {20}},\
  \bibinfo {pages} {204140} (\bibinfo {year} {2008})}\BibitemShut {NoStop}%
\bibitem [{\citenamefont {Alexander}\ \emph {et~al.}(2007)\citenamefont
  {Alexander}, \citenamefont {Bernoff}, \citenamefont {Mann}, \citenamefont
  {Mann}, \citenamefont {Wintersmith},\ and\ \citenamefont {Zou}}]{ABMMWZ07}%
  \BibitemOpen
  \bibfield  {author} {\bibinfo {author} {\bibfnamefont {J.~C.}\ \bibnamefont
  {Alexander}}, \bibinfo {author} {\bibfnamefont {A.~J.}\ \bibnamefont
  {Bernoff}}, \bibinfo {author} {\bibfnamefont {E.~K.}\ \bibnamefont {Mann}},
  \bibinfo {author} {\bibfnamefont {J.~A.}\ \bibnamefont {Mann}}, \bibinfo
  {author} {\bibfnamefont {J.~R.}\ \bibnamefont {Wintersmith}}, \ and\ \bibinfo
  {author} {\bibfnamefont {L.}~\bibnamefont {Zou}},\ }\href@noop {} {\bibfield
  {journal} {\bibinfo  {journal} {Journal of Fluid Mechanics}\ }\textbf
  {\bibinfo {volume} {571}},\ \bibinfo {pages} {191} (\bibinfo {year}
  {2007})}\BibitemShut {NoStop}%
\bibitem [{\citenamefont {Tucker}(2008)}]{T08}%
  \BibitemOpen
  \bibfield  {author} {\bibinfo {author} {\bibfnamefont {G.}~\bibnamefont
  {Tucker}},\ }\href@noop {} {\bibfield  {journal} {\bibinfo  {journal} {Harvey
  Mudd College Senior Theses}\ } (\bibinfo {year} {2008})}\BibitemShut
  {NoStop}%
\bibitem [{\citenamefont {de~Koker}\ and\ \citenamefont
  {McConnell}(1993)}]{KM93}%
  \BibitemOpen
  \bibfield  {author} {\bibinfo {author} {\bibfnamefont {R.}~\bibnamefont
  {de~Koker}}\ and\ \bibinfo {author} {\bibfnamefont {H.~M.}\ \bibnamefont
  {McConnell}},\ }\href {\doibase 10.1021/j100152a057} {\bibfield  {journal}
  {\bibinfo  {journal} {The Journal of Physical Chemistry}\ }\textbf {\bibinfo
  {volume} {97}},\ \bibinfo {pages} {13419} (\bibinfo {year} {1993})},\ \Eprint
  {http://arxiv.org/abs/http://pubs.acs.org/doi/pdf/10.1021/j100152a057}
  {http://pubs.acs.org/doi/pdf/10.1021/j100152a057} \BibitemShut {NoStop}%
\bibitem [{\citenamefont {Kent-Dobias}(2014)}]{K14}%
  \BibitemOpen
  \bibfield  {author} {\bibinfo {author} {\bibfnamefont {J.}~\bibnamefont
  {Kent-Dobias}},\ }\href@noop {} {\bibfield  {journal} {\bibinfo  {journal}
  {Harvey Mudd College Senior Theses}\ } (\bibinfo {year} {2014})}\BibitemShut
  {NoStop}%
\bibitem [{\citenamefont {Hansen}\ and\ \citenamefont {McDonald}(1990)}]{H90}%
  \BibitemOpen
  \bibfield  {author} {\bibinfo {author} {\bibfnamefont {J.-P.}\ \bibnamefont
  {Hansen}}\ and\ \bibinfo {author} {\bibfnamefont {I.~R.}\ \bibnamefont
  {McDonald}},\ }\href@noop {} {\emph {\bibinfo {title} {Theory of simple
  liquids}}}\ (\bibinfo  {publisher} {Elsevier},\ \bibinfo {year}
  {1990})\BibitemShut {NoStop}%
\bibitem [{\citenamefont {Bertsekas}(1999)}]{B99}%
  \BibitemOpen
  \bibfield  {author} {\bibinfo {author} {\bibfnamefont {D.~P.}\ \bibnamefont
  {Bertsekas}},\ }\href@noop {} {\emph {\bibinfo {title} {Nonlinear
  Programming}}},\ \bibinfo {edition} {2nd}\ ed.\ (\bibinfo  {publisher}
  {Athena Scientific},\ \bibinfo {address} {Belmont, MA},\ \bibinfo {year}
  {1999})\BibitemShut {NoStop}%
\end{thebibliography}%

\end{document}